\documentclass[superscriptaddress,floatfix,nofootinbib,preprintnumbers]{revtex4}

\usepackage{amsmath,graphicx,tabularx,url,times,fancybox}
\usepackage{footmisc,multirow}

\newcommand{\sgn}{\text{sgn}}
\newcommand{\gtot}{\Gamma^\text{tot}}
\newcommand{\gsm}{\Gamma^\text{SM}}
\newcommand{\ghid}{\Gamma^\text{hid}}
\newcommand{\gi}{\Gamma^\text{inv}}
\newcommand{\gv}{\Gamma^\text{vis}}
\newcommand{\ghh}{\Gamma^{HH}}
\newcommand{\ghhi}{\Gamma^{HH,\text{inv}}}
\newcommand{\ghhv}{\Gamma^{HH,\text{vis}}}
\newcommand{\bri}{\text{BR}^\text{inv}}
\newcommand{\brv}{\text{BR}^\text{vis}}

\newcommand{\brhhv}{\text{BR}^{HH,\text{vis}}}

\newcommand{\gev}{\text{GeV}}
\newcommand{\tev}{\text{TeV}}
\newcommand{\fb}{\text{fb}}

\newcommand{\ifb}{\text{fb}^{-1}}

\begin{document}

\title{Exploring the Higgs portal}

\author{Christoph Englert}
\affiliation{Institut f\"ur Theoretische Physik, 
             Universit\"at Heidelberg, Germany}

\author{Tilman Plehn}
\affiliation{Institut f\"ur Theoretische Physik, 
             Universit\"at Heidelberg, Germany}

\author{Dirk Zerwas}
\affiliation{LAL, IN2P3/CNRS, Orsay, France}

\author{Peter M. Zerwas} 
\affiliation{Deutsches Elektronen-Synchrotron
            DESY, Hamburg, Germany} 
\affiliation{Institut f\"ur Theoretische
            Teilchenphysik und Kosmologie, 
            RWTH Aachen University, Germany}

\date{August 19, 2011}

\begin{abstract}
  \noindent {\it We study the Higgs portal from the Standard-Model to
    a hidden sector and examine which elements of the extended theory
    can be discovered and explored at the LHC. Our model includes two
    Higgs bosons covering parameter regions where the LHC will be
    sensitive to two, one or none of the particles at typical
    discovery luminosities for Standard Model Higgs production. }
\end{abstract}

\maketitle

\section{Introduction}

Theoretical scenarios beyond the Standard Model [SM] which will be
tested at the LHC~\cite{bsm_review} often include a hidden sector. The
standard sector and the hidden sector are coupled by interactions of
gauge-invariant operators which open the gate for exploring structures
in the hidden sector by observing phenomena in the visible standard
sector. An attractive realization of this idea is provided by the
Higgs portal which connects the Higgs fields in the two sectors by an
elementary quartic interaction~\cite{hidhig1,higgs_portal,
  R1,wells,Schabinger:2005ei,R2,bij,Lebedev:2011aq,Chang:2006fp,
  Kanemura:2010sh,chacko,foot,hidvalley}.
Such a setup moves a precision study of the Higgs
sector~\cite{duehrssen,sfitter,Dedes:2008bf} into a central position of new physics
searches at the LHC.

Starting from a wide range of Higgs observables at the
LHC~\cite{higgs_review1,higgs_review2} its focus will naturally be on
measurements of Higgs masses, couplings and, to a lesser extent, Higgs
self-interactions particularly in cascade
decays~\cite{higgs_self}. The key observables which allow for such an
analysis are production rates for different decay channels combined
with the weak boson fusion process~\cite{wbf,vbfnlo} and the recently
revived associated production channels with decays to bottom
pairs~\cite{fatjet}.  For Higgs masses between 120 and 160~GeV the
LHC, running at a center-of-mass energy of 14~TeV and collecting
integrated luminosities in the $\mathcal{O}(10-100~\ifb)$ range, can
detect uncorrelated modifications to individual Higgs couplings of the
order of 30\% to 50\%~\cite{sfitter}.  Provided there exists some kind
of universal pattern in these modified couplings, the sensitivity
increases to 20\% or better~\cite{hidhig1}. A Higgs portal or hidden
Higgs sector is such a case with generally well-defined patterns in
the modified couplings. To render our analysis as transparent as
possible, we will illustrate the basic idea in a set-up in which
fields and interactions are isomorphic in the two sectors, just
supplemented by the quartic portal interaction. Adapting our results
to other models should be straightforward.\bigskip

We study the set of observables of a hidden Higgs sector and examine
to which extent it can be reconstructed by precision measurements in
collider experiments, {\it n.b.} at the LHC.  In an earlier, related
study~\cite{hidhig1} we restricted ourselves to the properties of the
SM-type Higgs boson.  To explore elements of the hidden sector,
invisible Higgs decays~\cite{eboli_zeppenfeld} to particles in this
sector play a crucial role.  Valuable additional insight we can obtain
from standard Higgs properties like masses and visible decay branching
ratios. The fundamental question whether a Higgs portal with
noticeable interactions between standard and hidden sector exists or
not, can be answered this way.

In this extended analysis we systematically explore the maximum
information that can be obtained on the Higgs portal and the
associated standard and hidden sector states from established Higgs
search strategies~\cite{prod12}, {\it i.e.} we consider both Higgs
masses lighter than ${\cal{O}}(1~\tev)$. For Higgs spectra with the
heavy narrow mass state in the trans-TeV region, analysis strategies
have been described in Ref.~\cite{wells}; discovery reaches for broad
and heavy states are discussed in, {\it e.g.},
Ref.~\cite{Bagger:1995mk}. The key question is how we can link the
parameters in our Higgs potential to general observables, like masses,
cross sections, or decay widths, and then to possible LHC
measurements, like twin width ratios~\cite{hidhig1}. While this work
is not meant to be an experimental analysis, realistically modelling
all statistical, systematic, and theoretical uncertainties, it defines
the strategy underlying such an analysis and points out its critical
steps from an experimental and theoretical point of view.

\section{From the potential to collider observables}
\label{sec:potobs}
Before we discuss realistic LHC prospects, it is important to study
the structure of Higgs portal models and identify the complete set of
observables which we can then try and access at the LHC.  The Higgs
potential we study in this letter consists of the Standard Model
component $[s]$, the isomorphic component in the hidden sector $[h]$,
and the quartic interaction coupling the two sectors with strength
$\eta_\chi$, {\it videlicet},
\begin{equation}
 \label{eq:potential}
 \mathcal{V} =
 \mu^2_s |\phi_s|^2 + \lambda_s |\phi_s|^4 
 \; + \;
 \mu^2_h |\phi_h|^2 + \lambda_h |\phi_h|^4
 \; + \;
 \eta_\chi |\phi_s|^2 |\phi_h|^2 \, .
\end{equation}
The mass parameters $\mu_j$ are generally substituted by $v_j$ after
expanding the two Higgs fields about their vacuum expectation values,
$\phi_j \rightarrow (v_j + H_j)/\sqrt{2}$ with $v_j^2 = (- \mu_j^2 -
\eta_\chi v_i^2 /2)/\lambda_j$ for $i \neq j = s,h$. The SM Higgs vacuum
expectation value is fixed by the gauge boson masses, since even in
the presence of a non-vanishing expectation value the hidden Higgs
fields do not contribute to electroweak symmetry breaking in the
standard sector. This is an important difference between the hidden
Higgs sector and other multi-Higgs models.  However, due to the
coupling of the two sectors the physical Higgs states in the SM and
the hidden sector mix to the mass eigenstates
\begin{alignat}{5}
  H_1 &=&  \cos\chi \, H_s + \sin\chi \, H_h &    \notag \\
  H_2 &=& -\sin\chi \, H_s + \cos\chi \, H_h & \; .
\label{eq:mixi}
\end{alignat}
Both, $H_1$ and $H_2$, couple to Standard Model fields through their
components $H_s$ and to the hidden sector through the admixtures
$H_h$.  For moderate coupling $\eta_\chi$ the properties of $H_1$
remain dominated by the Standard Model component, while the properties
of $H_2$ are characterized primarily by the hidden Higgs component.
The mixing of the fields in the potential generates self-interactions
among the light and heavy Higgs bosons~\cite{wells}, in particular
trilinear couplings $H_i H_j H_k$ of any combination. \bigskip

The phenomenology of the Higgs portal to the hidden sector depends on
the ratio of the Higgs boson masses. We will take $H_1$, primarily
$[s]$, to be the lighter particle and $H_2$, primarily $[h]$, to be
the heavier companion. Any scenario with other mass ratios could be
treated analogously but suffers from electroweak precision and
unitarity constraints.\bigskip

The properties of the two Higgs bosons are summarized in the masses,
$M_{1,2}$, the visible  and invisible widths, $\gv_{1,2}$
and $\gi_{1,2}$, both defined without including Higgs cascade decays,
and finally the Higgs cascade $\ghh_2$, realized by $H_2 \to H_1 H_1$
for suitable mass ratios. From these observables we can derive all
fundamental parameters of the Higgs potential.\bigskip

\noindent
{\it (i) \underline{Higgs masses}} --- 
Diagonalizing the Higgs mass matrix [squared],
\begin{equation}
  {\mathcal{M}}^2 = \left( \begin{array}{cc}
      2 \lambda_s v^2_s    &   \eta_\chi v_s v_h    \\
      \eta_\chi v_s v_h    &   2 \lambda_h v^2_h
    \end{array} \right)                             \,,
\end{equation}
generates the mass eigenvalues $M_{1,2}$ and the mixing angle
$\chi$,
\begin{alignat}{5}
 \label{eq:masses}
  M^2_{1,2} &= [\lambda_s v^2_s + \lambda_h v^2_h] \mp | \lambda_s
  v^2_s - \lambda_h v^2_h | \,
  \sqrt{1+\tan^2 2\chi} \\
  \tan \, 2\chi &= \frac{\eta_\chi v_s v_h}{\lambda_s v^2_s -
    \lambda_h v^2_h} \quad \text{with} \quad \pi/8 \leq \pm \chi \pm
  \pi/8 \leq 3\pi/8 \,,
\end{alignat}
for the two mass eigenstates $H_{1,2}$ defined in
Eq.~(\ref{eq:mixi}). The sign in front of $\chi$ coincides with
$\sgn[\eta_\chi]$ while the sign of the phase shift $\pm \pi/8$
corresponds to $\sgn[\lambda_s v_s^2 - \lambda_h v_h^2]$. The mixing
is restricted to
\begin{equation}
  \tan^2 2\chi \leq \dfrac{4 \lambda_s \lambda_h v^2_s v^2_h}
  {[\lambda_s v^2_s - \lambda_h v^2_h]^2}                           \,.
\end{equation}
For $v_h = v_s = 246$ GeV and $\lambda_s = \lambda_h /4 = 1/8$, a
parameter set reminiscent of the Standard Model, we illustrate the two
Higgs masses as functions of the mixing parameter in
Fig.~\ref{fig:Masses12}. They are compared with the bounds derived
from the non-observation of Higgs bosons at LEP for standard and
reduced couplings~\cite{LEPred}. For this illustrational parameter set
the mixing has to stay below $\sin\chi \leq 0.22$, as a direct result
of the LEP bound on the Higgs mass. This kind of bound is a general
feature, because the mixture of a Standard Model and a relatively
light hidden Higgs state will generate one mass eigenvalue below the
SM diagonal entry in Eq.~(\ref{eq:masses}).\bigskip

Unitarity for high energies and precision observables like the $\rho$
parameter constrain the mass values in complete analogy to the
Standard Model case. The usual SM Higgs mass or its logarithm is
substituted by the superposition of the two Higgs masses, weighted by
the mixing parameters $\cos^2 \chi$ and $\sin^2 \chi$, {\it e.g.},
\begin{alignat}{5}
\label{eq:unitarity}
&\text{unitarity} &M^2_{H_\text{SM}} &\to \langle M_i^2 \rangle \equiv
\cos^2 \chi\, M^2_1 + \sin^2 \chi\, M^2_2 \leq
4 \pi \sqrt{2} / 3 G_F \simeq (700 \,\gev)^2            \notag \\
&\text{$\rho$ parameter} \quad &\log{M^2_{H_\text{SM}}} &\to \langle
\log{M_i^2} \rangle \equiv \cos^2 \chi \log M^2_1 + \sin^2 \chi
\log{M^2_2} \leq \log (175 \,\gev)^2 \,.
\end{alignat}
As expected, for small mixing the SM bounds transfer to $M_1$ while
$M_2$ remains essentially unconstrained. However, for large mixing the
two bounds transfer to the algebraic and geometric means of the $M_1,
M_2$ mass pair, thus reducing the allowed range for $M_2$
considerably. Finally, for large $\sin^2 \chi$, $M_1$ and $M_2$
interchange their roles.

Because the most restrictive bounds arise from electroweak precision
data we base our numerical scan over the Higgs potential on the
complete set of $S,T,U$ parameters~\cite{Peskin:1990zt}.  Confronting
the model defined in Eq.~(\ref{eq:potential}) with the current bounds
on $S,T,U$~\cite{Alcaraz:2006mx}, we need to emphasize one caveat: if
the spontaneous symmetry breaking in the hidden sector also gives rise
to additional massive states, the $S,T,U$
values~\cite{wells,Peskin:2001rw,Ahlers:2008qc} may be altered
significantly, {\it e.g.}, through kinetic $U(1)$
mixing~\cite{Holdom:1985ag}. Since we are primarily interested in the
phenomenology of the two Higgs states, Eq.~(\ref{eq:masses}), we do
not further investigate this direction, yet we also consider parameter
points of the portal model which violate the bounds by about
$10\%$.\bigskip

\begin{figure}[!t]
\includegraphics[height=0.30\textwidth]{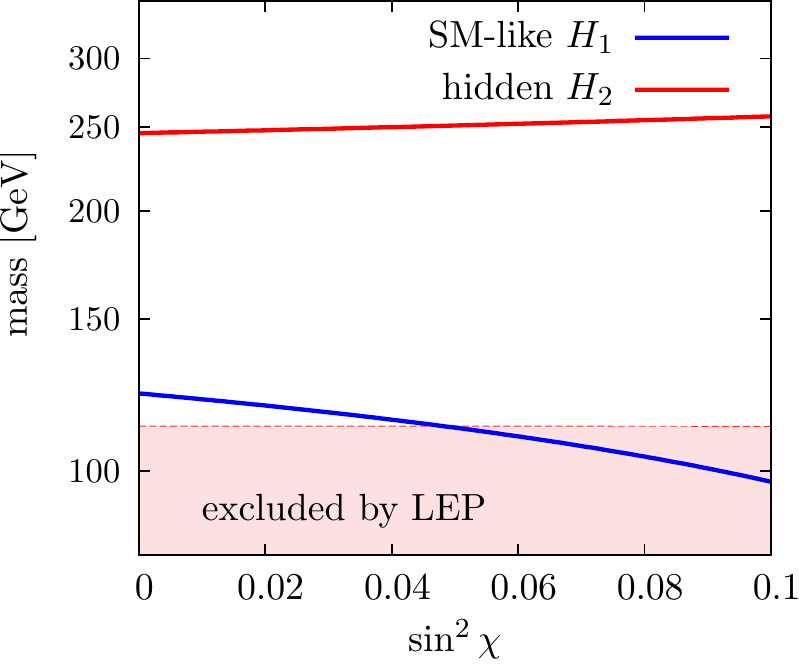}
\hspace*{0.15\textwidth}
\includegraphics[height=0.30\textwidth]{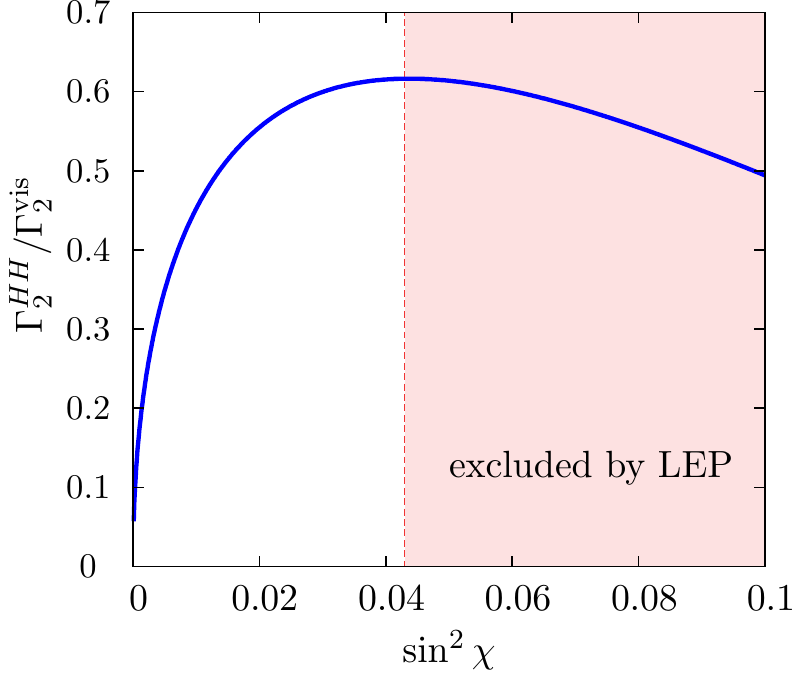}
\caption{\label{fig:Masses12} \it Left: masses of the light SM-like
  Higgs boson $H_1$ (blue) and the heavy Higgs boson $H_2$ (red). The
  parameters of the Higgs potentials are $v_h = v_s = 246$ GeV and
  $\lambda_s = \lambda_h /4 = 1/8$.  The shaded region displays the
  LEP bound~\cite{LEPred}. Right: cascade decay width $\ghh_2$ as a
  function of $\sin^2\chi\,$ for the same parameters. Again, the
  region in which $H_1$ is excluded by LEP is shaded.}
\end{figure}

\noindent {\it (ii) \underline{Direct Higgs production and decay}} ---
All $H_1$ couplings to SM-particles are suppressed by the mixing
parameter $\cos\chi$, $H_2$ couplings correspondingly by $\sin\chi$;
for direct decays to particles in the hidden sector the two
coefficients are reversed:
\begin{alignat}{5}
  \gv_1  =& \cos^2\chi \, \gsm_1 \qquad \text{and} \qquad 
& \gv_2  =& \sin^2\chi \, \gsm_2                            \notag \\
  \gi_1  =& \sin^2\chi \, \ghid_1 \qquad\; \text{and} \qquad
& \gi_2  =& \cos^2\chi \, \ghid_2                           \,.  
\end{alignat}
While $\gsm_{1,2}$ denote the SM Higgs widths for the eigenstates with
masses $M_{1,2}$, $\ghid_{1,2}$ play the same role in the hidden
sector, {\it i.e.} Higgs cascade decays $H_2 \to H_1 H_1$ not
included. The same index `vis' may also specify a subset of final
states.  The cross sections for Higgs production channels in all
SM-particle collisions are affected analogously
\begin{equation}
  \sigma_{1}  = \cos^2\chi \, \sigma^\text{SM}_1 \qquad \text{and} \qquad
  \sigma_{2}  = \sin^2\chi \, \sigma^\text{SM}_2                               \,.
\end{equation}
At hadron colliders, we measure a particular combination of partial
widths and production cross sections as the twin-width
ratios~\cite{hidhig1}
\begin{equation}
  \kappa_j = 
  \dfrac{\left(\dfrac{\Gamma^p \Gamma^d}{\gtot}\right)_j}
        {\left(\dfrac{\Gamma^p \Gamma^d}{\gtot} \right)^\text{SM}_j}
  \quad \text{for} \quad j=1,2                                              \,.
\label{eq:twin}
\end{equation}
The partial widths refer to the production channel $p$ and the decay
mode $d$, either exclusively or summed over sets of initial or final
states.  These ratios are measured, at the Born level, directly by the
product of production cross section times decay branching ratio of the
process $p \to Higgs \to d$ in the narrow-width approximation.  The
parameters $\kappa_{1,2}$ are independent of $p,d$ with values less
than unity, and any pair $[p,d]$ may be exploited to measure
$\kappa_j$. The universality of the parameters $\kappa_{1,2}$ is a
signal for mixed Higgs states or a universal coupling modification due
to a naive bound state form factor~\cite{Giudice:2007fh,hidhig1}.
Joining the measurement of the $\kappa_j$ with the invisible branching
ratios allows us to determine the mixing parameter $\sin^2
\chi$. \bigskip

\noindent {\it (iii) \underline{Invisible Higgs decay}} --- For $H_1$
the invisible decay width is generated merely by direct decays into
the hidden sector, for $H_2$ the direct decays may be supplemented by
cascades.  For this argument we assume that the standard Higgs decays
$H \to ZZ \to 4\nu$ are subtracted after measuring the corresponding
$4\ell$ modes. 
For light states $H_1$ there is a non-negligible decay rate of the 
Higgs into gluons and charm quarks. These branching ratios are 
phenomenologically inaccessible at the LHC due to large backgrounds. 
This adds to complications in extracting the Higgs-to-invisible 
branching ratio, which are already familiar from the SM scenario. Two methods,
consistent with the present portal scenario, can be followed to solve the problem. 
Either these modes may be included theoretically into the {\it corpus} of visible 
final states by adopting the SM couplings, as assumed in the model. 
Or the issue can be treated consistently by fitting a partial width 
for the undetected decays as part of a global fit to the model 
as described in Refs.~\cite{duehrssen,sfitter}.
In relation to the visible decay widths the invisible
decay widths can be extracted by measuring the ratios of the
corresponding branching ratios, {\it i.e.}
\begin{alignat}{5}
  \frac{\gi_1}{\gv_1} &= \frac{\bri_1}{\brv_1}              \notag       \,, \\
    \frac{\gi_2 + \ghhi_2}{\gv_2} &= \frac{\bri_2}{\brv_2} \,,
\end{alignat}
where $\ghhi$ denotes the invisible decay width resulting from the
cascade decay $H_2\rightarrow H_1H_1$ with invisible $H_1$ decay
modes. In contrast to the {\it ad-hoc} definition of the visible
branching ratio $\brv_2 = \gv_2/\gtot_2$, the measured invisible
branching ratio $\bri_2$ necessarily includes the invisible cascade
decays.\bigskip

\noindent {\it (iv) \underline{Higgs cascade}} --- If $H_2$ is
sufficiently heavier than $H_1$, the cascade channel $H_2 \to H_1 H_1$
opens up with its partial width
\begin{equation}
\label{eq:cascadedec}
\ghh_2 = \frac{\Lambda^2_{211}}{32 \pi} \, \frac{\beta_1}{M_2}      \,.
\end{equation}
The velocity of $H_1$ in the rest frame of $H_2$ is denoted by
$\beta_1$ while the effective $H_2 H_1 H_1$ coupling, derived by
inserting the mixed states into the potential $\mathcal{V}$, reads
\begin{alignat}{5}
  \label{eq:lambda211}
  \Lambda_{211} =& \; 3 \sin 2\chi \left[ \cos\chi \, \frac{\lambda_s
      v^2_s}{v_s}
    -\sin\chi \, \frac{\lambda_h v^2_h}{v_h} \right] \notag \\
  & - \tan 2\chi \, [\lambda_s v^2_s - \lambda_h v^2_h] \, \left[
    (1-3\cos^2\chi) \, \frac{\sin\chi}{v_h} -(1-3\sin^2\chi) \,
    \frac{\cos\chi}{v_s} \right] \,.
\end{alignat}
The decays of the $H_1 H_1$ pair give rise to visible-visible,
visible-invisible, and invisible-invisible final states with
probabilities $\cos^4\chi$, $2 \sin^2 \chi \cos^2 \chi$ and
$\sin^4\chi$, respectively.  As a result, we can reconstruct $\ghh_2$
from the channel in which both $H_1$ decays are visible: $\ghh_2 =
\ghhv_2 / \cos^4 \chi$. To illustrate the probability of cascade
decays we show the $\chi$-dependence of the ratio $\ghh_2 / \gv_2$ in
the right panel of Fig.~\ref{fig:Masses12}.\bigskip

From all observables listed above we can derive the fundamental
properties of the two Higgs bosons, which are related to the dynamics
in the hidden sector: the mixing angle $\sin\chi$, the invisible
partial widths $\gi_{1,2}$, the cascade width $\ghh_2$ and the total
widths $\gtot_{1,2}$. The latter are notorious at hadron
colliders. While we cannot experimentally determine them for narrow
states, they are crucial properties of our two Higgs states.  Provided
$H_2$ is heavier than twice the $H_1$ mass, the total widths of the
two Higgs bosons are given in terms of five partial widths
\begin{alignat}{5}
  \gtot_1 &= \cos^2 \chi \, \gsm_1 + \sin^2 \chi \, \ghid_1      \notag \\
  \gtot_2 &= \sin^2 \chi \, \gsm_2 + \cos^2 \chi \, \ghid_2 + \ghh_2
  \,.
\end{alignat}
Ratios of partial and total Higgs widths, however, are observable at
the LHC. For the light SM-type Higgs boson the relations
\begin{alignat}{5}
  \dfrac{\gi_1}{\gsm_1} &= \cos^2\chi \, \left[
    \dfrac{\cos^2\chi}{\kappa_1} -1 \right] \,, \qquad\quad
  & \dfrac{\ghh_1}{\gsm_1} &= 0  \,,                                                  \notag \\
  \dfrac{\gi_1}{\gsm_1} &= \cos^2\chi \, \dfrac{\bri_1}{\brv_1} \,,
  \;\qquad\qquad\quad\;\; & \dfrac{\gtot_1}{\gsm_1} &=
  \dfrac{\cos^4\chi}{\kappa_1} \,.
\label{eq:w1}
\end{alignat}
link the decay width to SM particles $\gsm_1$ to the total width
$\gtot$ and provide us with two equivalent expressions for the
modified invisible branching ratio. Thus, we can express the mixing
angle $\cos^2 \chi$ and $\gi_1/\gsm_1$ in terms of the observable
twin-width ratio $\kappa_1$ and the branching ratios $\bri_1/\brv_1$.

The corresponding expressions for the heavy hidden-type Higgs boson
$H_2$ are slightly modified because they include the cascade decay,
followed by the decay of the $H_1$ pair back to the visible sector
\begin{alignat}{5}
  \dfrac{\gi_2}{\gsm_2} &= \sin^2\chi \, \left[
    \dfrac{\sin^2\chi}{\kappa_2} -1 - \dfrac{1}{\cos^4\chi} \,
    \dfrac{\brhhv_2} {\brv_2} \right] \,, \qquad\quad
  &\dfrac{\ghh_2}{\gsm_2} &= \dfrac{\sin^2\chi}{\cos^4\chi} \,
  \dfrac{\brhhv_2}
  {\brv_2} \,,                                \notag \\
  \dfrac{\gi_2}{\gsm_2} &= \sin^2\chi \, \left[ \dfrac{\bri_2}{\brv_2}
    \, - \tan^4\chi \, \dfrac{\brhhv_2} {\brv_2} \right] \,, &
  \dfrac{\gtot_2}{\gsm_2} &= \dfrac{\sin^4\chi}{\kappa_2} \,.
\label{eq:w2}
\end{alignat}
The mixing is restricted by the positivity of the decay widths to the
band
\begin{equation}
\label{eq:limkappa}
  \kappa_2 \leq \sin^2 \chi \leq 1 - \kappa_1  \; .
\end{equation}
%
 
\begin{figure}[!t]
\includegraphics[height=0.30\textwidth]{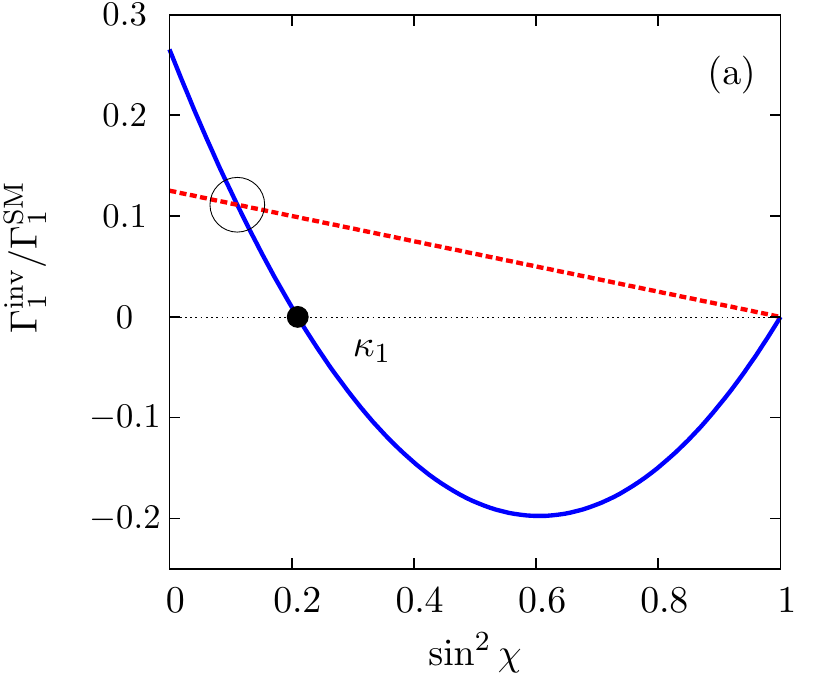}
\hspace*{0.15\textwidth}
\includegraphics[height=0.30\textwidth]{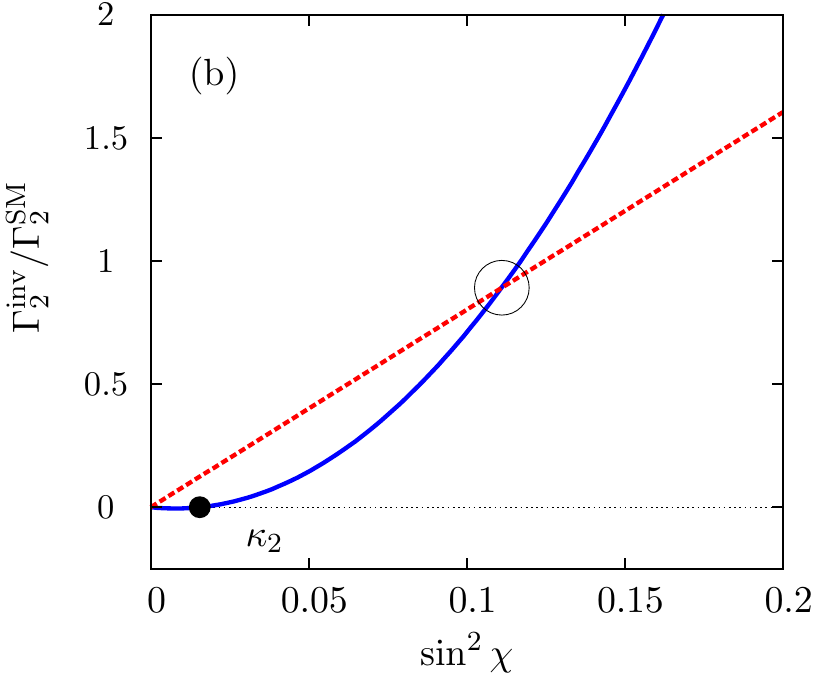}
\caption{\label{fig:kagi} \it Illustration of the intersections of the
  solutions for $\kappa_j$ (blue) and the invisible branching ratios
  (red), generating the physical solutions for the mixing parameter
  and the invisible widths. The Higgs portal model parameters are
  chosen $v_s=246~\text{GeV}$, $v_h=v_s$, $\lambda_s=1/8$ and
  $\lambda_h=4\lambda_s$; the intersection points of the curves
  correspond to the solution $\sin\chi =1/3$ for the mixing
  parameter.}
\end{figure}
 
\begin{table}[!b]
\begin{tabular}{c|c|c}
\hline
                        & parameters                                & measurements                                                  \\
\hline 
\multirow{2}{*}{Higgs properties}        & $M_{1,2}$                                 & resonance peaks                                         \\
                        & $\sin^2\chi$, $\gi_{1,2}$                 & $\kappa_{1,2}$, $\langle \sigma \bri \rangle$   \\
\hline
\multirow{5}{*}{Higgs potential}  & $\lambda_s v_s^2$, $\lambda_h v_h^2$,
                                             $\eta_\chi v_s v_h \;\;$     & $M_{1,2}$, $\sin^2\chi$                                 \\ 
                        & $v_s$                                           & $M_W$                                                   \\
                        & $v_h$                                           & $\Gamma_2^{HH}$                                         \\[2mm]
                        & $\Downarrow$                                   &                                                         \\[2mm]
                        & $\lambda_s$, $\lambda_h$, $\eta_\chi$           & {\it derived subsequently}                              \\
\hline
\end{tabular}
\caption{\label{tab:params} Consecutive determination of the Higgs
  properties and parameters of the Higgs portal potential.}
\end{table}

In Fig.~{\ref{fig:kagi}} we show the numerical solutions to
Eqs.(\ref{eq:w1}) and (\ref{eq:w2}).  They return unique values of the
mixing parameter $\sin^2\chi$ and the partial widths $\gi_{1,2}$ for
decays of the two Higgs bosons $H_{1,2}$ to particles in the hidden
sector.  The parameters chosen in these figures are the same as for
the Higgs masses in Fig.~{\ref{fig:kagi}}, supplemented for
illustration by the identities $\ghid_{1,2} \equiv
\gsm_{1,2}$.\bigskip

Finalizing our argument, the set of phenomenological Higgs parameters
is indeed sufficient for the reconstruction of the parameters in the
Lagrangian:
\begin{equation}
  \{M_{1,2}, \sin \chi, \Gamma^{HH}_2; M_W \}
  \qquad \Longleftrightarrow \qquad
  \{v_s, v_h; \lambda_s, \lambda_h; \eta_\chi \} \,,
  \notag
\end{equation}
where further details we give in Tab.~\ref{tab:params}.  The two
values $\lambda_s v^2_s$, $\lambda_h v^2_h$ and $\eta_\chi v_s v_h$
[{\it modulo} the discrete sign ambiguity] can be derived from the
measured Higgs boson masses $M_{1,2}$ and the mixing parameter $\tan^2
2 \chi$.  The vacuum expectation value $v_s$ is directly related to
the $W$ boson mass $M_W = gv_s/2$, while the only remaining free
parameter $v_h$ is fixed by measuring the magnitude of the Higgs
self-coupling $\Lambda_{211}$.  A quantitative phenomenological
example for reconstructing the theory this way we discuss in the
following section.


\section{Higgs profiling at the LHC}
\label{sec:hprofile}
The phenomenological analysis of the hidden-Higgs theory, after the
discovery of the Higgs boson(s), splits into two parts. First, we
determine the parameter regions where two, one or none of the Higgs
particles can be detected. Subsequently, we study to which extent we
can reconstruct the parameters of the standard and hidden Higgs
sectors at the LHC.\bigskip

The analyses will be exemplified by choosing the same value for $\gsm$
as for the invisible width $\ghid$, which encodes the dynamics of the
hidden sector.  To asses the sensitivity of the LHC experiments at a
typical SM Higgs discovery luminosity ${\cal{L}}=30~\ifb$ we randomly
scan over the (hidden) Higgs potential of
Eq.~(\ref{eq:potential}). Statistical significances we derive by
rescaling the results of the experimental simulations in
Ref.~\cite{prod12} for parameter choices consistent with the
constraints from unitarity and electroweak precision
data~\cite{Alcaraz:2006mx}.  We choose $(S,T)=(0,0)$ for
$m_t=170.9~\gev$~\cite{cdf:2007bxa} and $m_H^\text{SM}=115~\gev$. We
have verified our implementation against Ref.~\cite{wells}.\bigskip

First, we identify the part of parameter space where at least one
Higgs boson can be discovered at the LHC in a visible channel. In
terms of the masses $M_{1,2}$ and the mixing parameter $\sin^2 \chi$
we show this area in Fig.~\ref{fig:Mx}(a,b). The light SM-type Higgs
boson is clearly visible for small $\sin^2 \chi$, the hidden-type
Higgs boson for large $\sin^2 \chi$. In Fig.~\ref{fig:Mx}(c,d) 
the parameter space of $M_1,M_2,\sin^2\chi$ is shown in which both Higgs
bosons can be detected at the LHC. Combining them, we can distinguish
four different areas for $\sin^2 \chi$ where either none or one or two
Higgs bosons are accessible at the LHC for $\mathcal{L} = 30~\ifb$:
\begin{center} 
\begin{tabular}{c|crl}
  \hline
  & \# observable Higgs bosons$\;\;$  &   & \\
  \hline
  $\sin^2 \chi\lesssim 0.2$        & 1  & SM-type Higgs $H_1\;$  & ($\sigma_{H_1}\geq 3,~\sigma_{H_2}\leq 1$)        \\
  $0.3 \lesssim \sin^2 \chi \lesssim 0.4$  & 0  &  neither SM-type nor hidden-type Higgs $H_1,H_2\;$  & ($\sigma_{H_1,H_2}< 3$) \\
  $ 0.4\lesssim \sin^2 \chi \lesssim 0.6$  & 2  & SM + hidden-type Higgs $H_1,H_2\;$ &  ($\sigma_{H_1,H_2}\geq 3$) \\
  $\sin^2\chi \gtrsim 0.6$          & 1  & hidden-type Higgs $H_2\;$ & ($\sigma_{H_1}\leq 1,~\sigma_{H_2}\geq 3$)      \\ 
  \hline
\end{tabular} \end{center}
When we steadily increase the hidden widths the fraction of parameter
space in which none of the Higgs bosons can be detected
increases. Thus, overwhelming decay modes into the hidden sector can
screen Higgs bosons in a natural way [and non-observation of Higgs
bosons, for limited luminosity, does not necessarily imply the
falsification of the Higgs mechanism].\bigskip

\begin{figure}[!t]
    \includegraphics[height=0.35\textwidth]{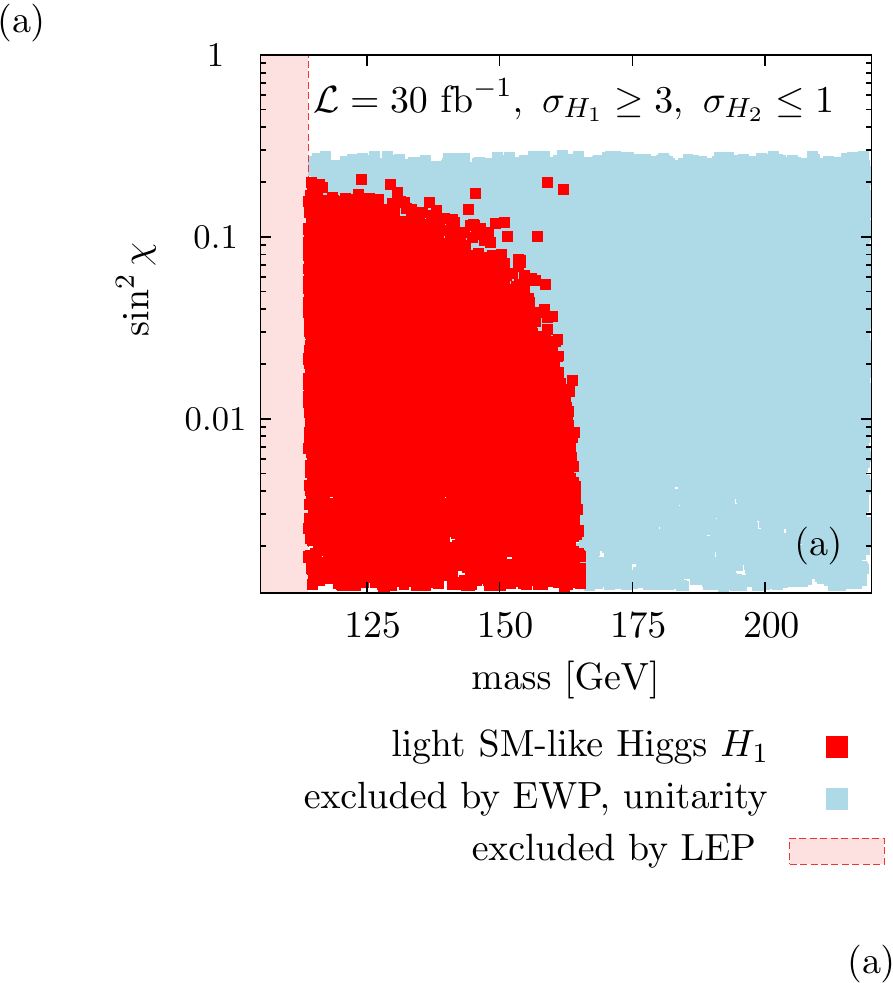}
    \hspace*{0.1\textwidth}
    \includegraphics[height=0.35\textwidth]{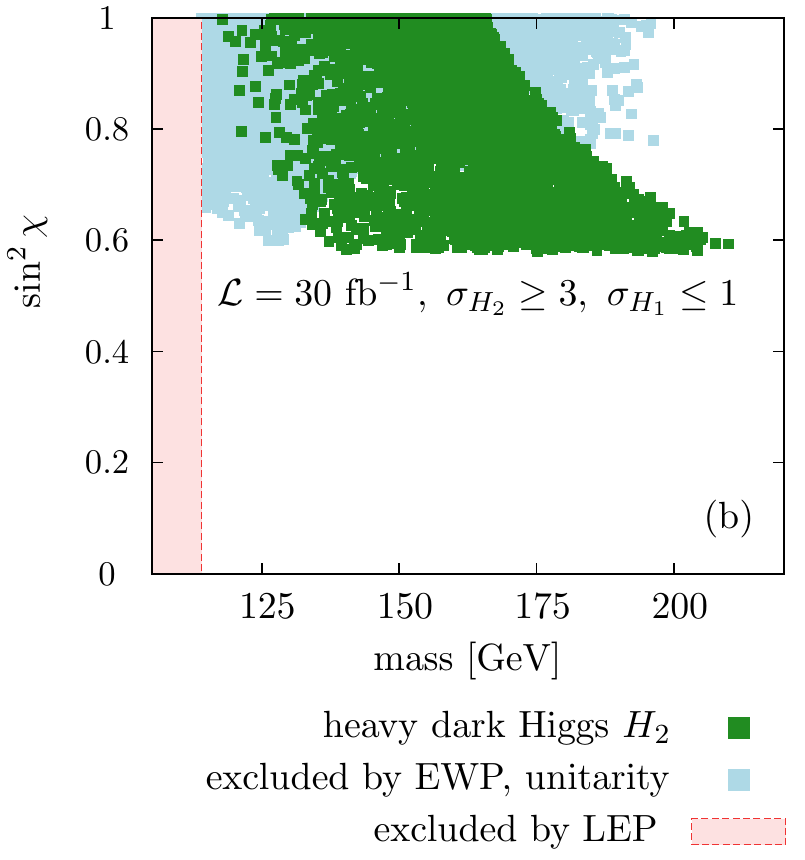} \\[4mm]
    \includegraphics[height=0.285\textwidth]{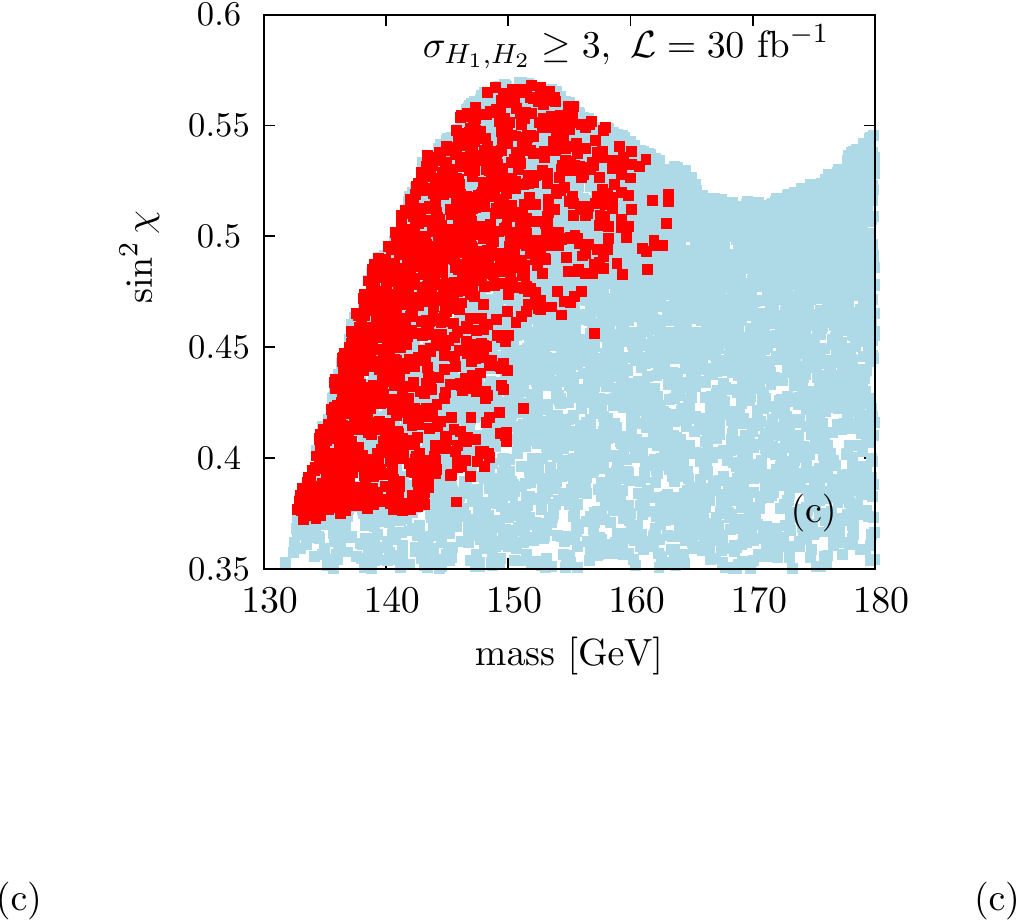}
    \hspace*{0.08\textwidth}
    \includegraphics[height=0.285\textwidth]{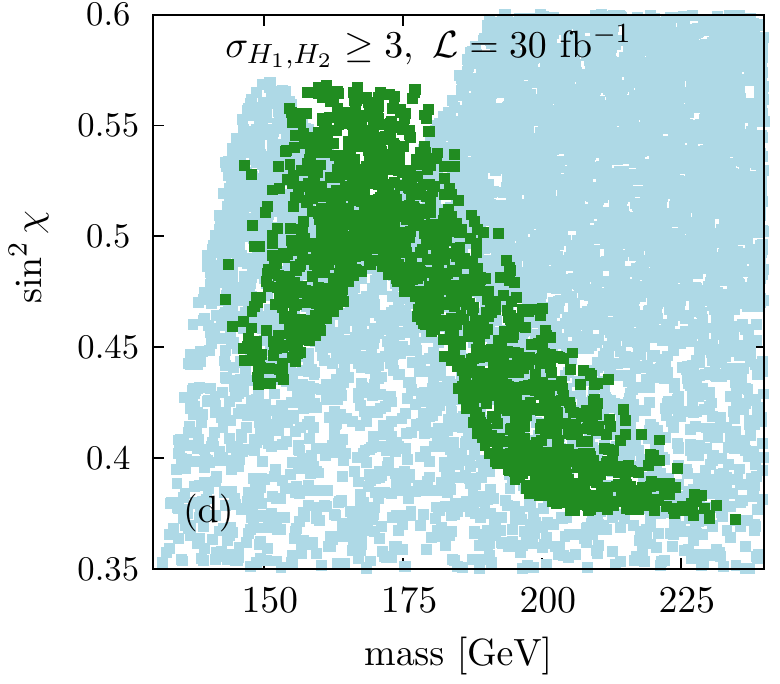}
    \caption{\label{fig:Mx} \it Randomly generated Higgs portal
      models. The parameters ranges are
      $v_h\in(0~\text{GeV},246~\text{GeV}], v_s= 246~\text{GeV},
      \lambda_h\in(0,4\pi], \lambda_s\in(0,4\pi]$, and
      $\eta_\chi\in[-4\pi,4\pi]$. The hidden Higgs decay width we
      identify with the SM decay width, {\it i.e.}  $\ghid \equiv
      \gsm$. LEP constraints and bounds from
      $S,T,U$~\cite{Alcaraz:2006mx} are included, likewise the
      unitarity constraint of Eq.~(\ref{eq:unitarity}). Panel~(a)
      displays the sensitivity for $H_1$ only, panel~(b) for $H_2$
      only, and panels~(c) and (d) show where the LHC is sensitive to
      both $H_1$ and $H_2$ at the same time for $30~\ifb$. Cross
      sections and widths we compute using {\sc Higlu}~\cite{higlu}
      and {\sc Hdecay}~\cite{hdecay}.}
\end{figure}

By measuring the production of the two Higgs bosons in the visible and
invisible decay channels, we can determine the properties of the Higgs
bosons, as demonstrated in the previous section: the masses $M_{1,2}$;
the mixing parameter $\sin^2\chi$; and the invisible widths
$\gi_{1,2}$. Since invisible channels are involved, the full capacity
of LHC with luminosity of $300~\ifb$ will be needed to draw a
finely-grained picture. The analysis would be redundant if the two
Higgs bosons had been distinguishable in the invisible channel, which
turns out to be a very difficult task.\footnote{Varying the LHC energy
  from 7 to 14 TeV, the $H_1$ and $H_2$ components may in principle be
  disentangled. Missing mass techniques in $e^+ e^-$ collisions, {\it
    cf.} Ref.~\cite{LC}, allow for a separation of the two Higgs
  bosons also in invisible decays.}

Fortunately, the parameters can also be determined if only the
superposition of the two Higgs bosons is measured in the invisible
channels. We define the weighted ratio of invisible over visible
branching ratios by
\begin{equation}
\left\langle \dfrac{\sigma\bri}{\sigma\brv} \right\rangle = 
\dfrac{\langle \sigma_1 \bri_1 + \sigma_2 \bri_2 + \text{Bkg} \rangle_\text{inv}}
      {\langle \sigma_1 \brv_1 + \sigma_2 \brv_2 + \text{Bkg} \rangle_\text{vis}} \,,
\end{equation}
where the subscripts 'inv' and 'vis' refer to different analysis
strategies in measuring visible and invisible decays~\cite{prod12}.
The production cross sections and branching ratios of the signal we
can re-express in terms of the $\kappa_{1,2}$ parameters and the
mixing angle:
\begin{alignat}{5}
  \sigma_j \bri_j =& \sigma^\text{SM}_j \left[ \left\{ \begin{matrix}
     \cos^2\chi \\ \sin^2\chi \end{matrix}\right\} - \kappa_j \left(1 + h^\text{inv}_{j=2}\right)
  \right]
  \quad \text{for} \quad j = 1,2                                                  \notag \\
  \sigma_j \brv_j =& \sigma^\text{SM}_j \kappa_j \,.
\end{alignat}
The $h$ term denotes generally small complements from $H_2 \to H_1
H_1$ cascade decays, which can be measured in the visible channels:
$h^\text{inv}_2 = - \kappa_2 [\brhhv_2 /\brv_2] \, (1+\sin^2
\chi)/\cos^2 \chi$. Since the invisible and visible channels are
treated incoherently, the measurement errors can be slightly reduced
by restricting the analysis to the invisible decay modes $\langle
\sigma \bri \rangle$ instead of the ratio, {\it cf.}
Fig.~\ref{fig:avBR}.\bigskip

We adopt the experimental simulations of Ref.~\cite{prod12} for both
Higgs bosons. Since the $\chi$ dependence of the cross section for
invisible decays of the two-Higgs system is predicted to be
\begin{equation}
  \label{eq:disentangle}
  \langle \sigma \bri \rangle \sim const - [\cos^2 \chi + 
  \{\sigma^\text{SM}_2 / \sigma^\text{SM}_1\} \, \sin^2 \chi] \, ,
\end{equation}
we lose all sensitivity if the two Higgs masses are too close and the
ratio of the two SM cross sections is near unity. However, for
increasing $H_2$ masses, the $H_2$ contribution steadily drops with
the production cross section.  This leads to the asymptotic behavior
$\langle \sigma \bri \rangle \sim \text{const} - \cos^2 \chi$
generated by $H_1$ alone; potential errors induced by the
extrapolation in the $H_2$ mass are suppressed as a result.

The cross section $\langle \sigma \bri \rangle$ for invisible decays
we illustrate in Fig.~{\ref{fig:avBR}} for three reference points in
which invisible and visible decays of the two Higgs bosons are nearly
balanced. The reference points are characterized by the $H_1, H_2$
masses and the true mixing angle:
\begin{alignat}{5}
 \label{eq:reference}
    \text{ref. point \#1~}\quad\{M_1, M_2; \sin^2 \chi\} &= \{140, 160; 0.456\} \,, \notag \\
    \text{ref. point \#2~}\quad\{M_1, M_2; \sin^2 \chi\} &=\{115, 300;0.25\} \,, \\
    \text{ref. point \#3~}\quad\{M_1, M_2; \sin^2 \chi\} &=\{115, 400; 0.25\}\,, \notag 
\end{alignat}
with masses in units of GeV.  
For these parameter points the discovery of the Higgs bosons
  is steered by the visible decay channels $H_1\rightarrow b\bar
  b,\gamma\gamma$ ($M_1=115~\gev$) and $H_i\rightarrow 4\ell$
  ($M_1=140~\gev, M_2=300, 400~\gev$).
Ref.$\!$ point \#3 falls mildly outside
the 95\% confidence level $S,T$ contour: $S=0.018,~T=-0.031 <
T_{95\%}=-0.028$. It exemplifies the asymptotic behavior of a large
hierarchy between the two production rates $\sigma_2^\text{SM} /
\sigma_1^\text{SM} \sim 3~\fb / 44~\fb$.  The production rates include
gluon fusion~\cite{gfusion} as cast in {\sc Higlu}~\cite{higlu} and
electroweak vector boson fusion {\sc Vbfnlo}~\cite{vbfnlo}; branching
ratios for decay channels we adopt from {\sc
  Hdecay}~\cite{hdecay}. Background estimates we incorporate by
adjusting the simulations described in Ref.~\cite{prod12}. From the
three measurement shown, the value of $\sin^2\chi$ can be determined
uniquely.  Given this mixing parameter, we can finally calculate the
partial widths for invisible Higgs decays into the hidden sector, as
shown in Eqs.~({\ref{eq:w1}) and (\ref{eq:w2}}).  The result,
including only statistical effects on the extraction of $\sin^2\chi$
we show in Tab.~\ref{tab:sinex}.

\begin{table}[!b]
  \begin{tabular}{l|c|c|c}
    \hline
    & reference point \#1        & reference point \#2     & reference point \#3        \\
    \hline
    $\sin^2\chi$          &  $0.46\pm 0.25$   &   $0.24\pm 0.04$ ($0.25\pm 0.03$) &  $0.22\pm 0.15$ ($0.24\pm 0.10$)     \\ 
    $\gi_1 / \gsm_1$   &  $0.46\pm 0.67$   &  $0.25\pm 0.07$ ($0.25\pm 0.05$)  & $0.29\pm  0.26$ ($0.27\pm 0.17$)     \\ 
    $\gi_2 / \gsm_2$   &   $0.54\pm 0.83$  &  $2.86\pm 0.74$ ($2.89 \pm 0.54$) &  $0.01\pm  0.29$ ($0.02\pm 0.23$)     \\
    \hline
  \end{tabular}
  \caption{\label{tab:sinex} \it
    Parameters extracted for the three reference points defined in Eq.~(\ref{eq:reference}); 
    the luminosity is set to $\mathcal{L} = 300~\ifb~(600~\ifb)$. For reference point \#1
    increased statistics does not amount to smaller errors since the
    two states $H_1,H_2$ are too close in mass.}
\end{table}

We see that depending on the specific realization of the Higgs portal
we can reconstruct some, if not all, elements of the underlying
theory, {\it i.e.}  the vacuum expectation values and quartic
couplings in the standard and hidden sectors, as well as the coupling
of the Higgs portal.  If the invisible decays of the heavier state
$H_2$ can be decoupled from $\left\langle \sigma\bri\right\rangle$ by
suppressing the production cross section, Eq.~(\ref{eq:disentangle}),
the value of $\sin^2\chi$ can eventually be inferred from the
invisible Higgs decays.  The measurement of the self-coupling
$\Lambda_{211}$ provides additional information, which further
constrains the allowed parameters of the hidden sector. After
$\lambda_h v^2_h$ is determined by measuring the Higgs masses and the
mixing parameter, the coupling $\lambda_h$ and the vacuum expectation
value $v_h$ can be disentangled in cascade decays.\bigskip

\begin{figure}[!t]
      \includegraphics[height=0.30\textwidth]{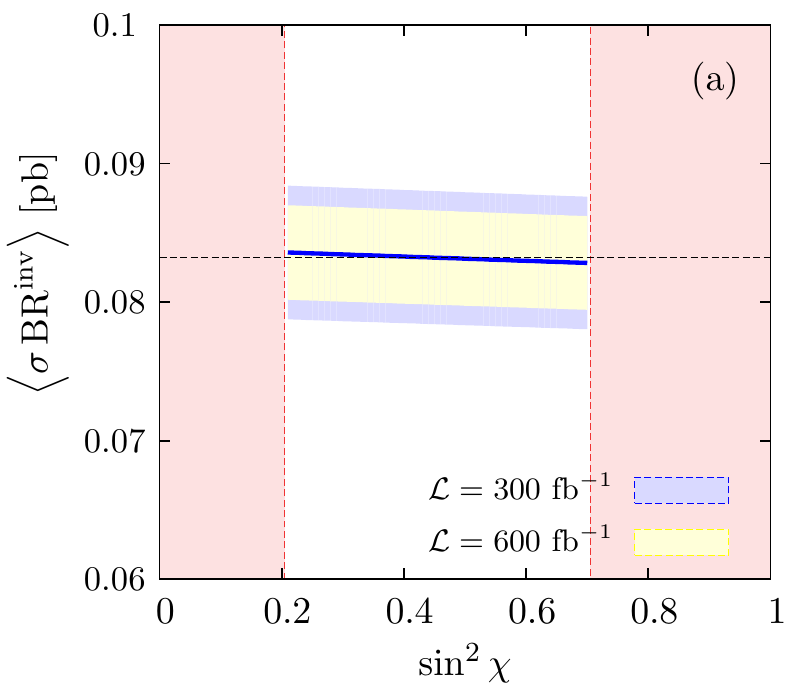} 
      \hspace{0.1\textwidth}
    \includegraphics[height=0.30\textwidth]{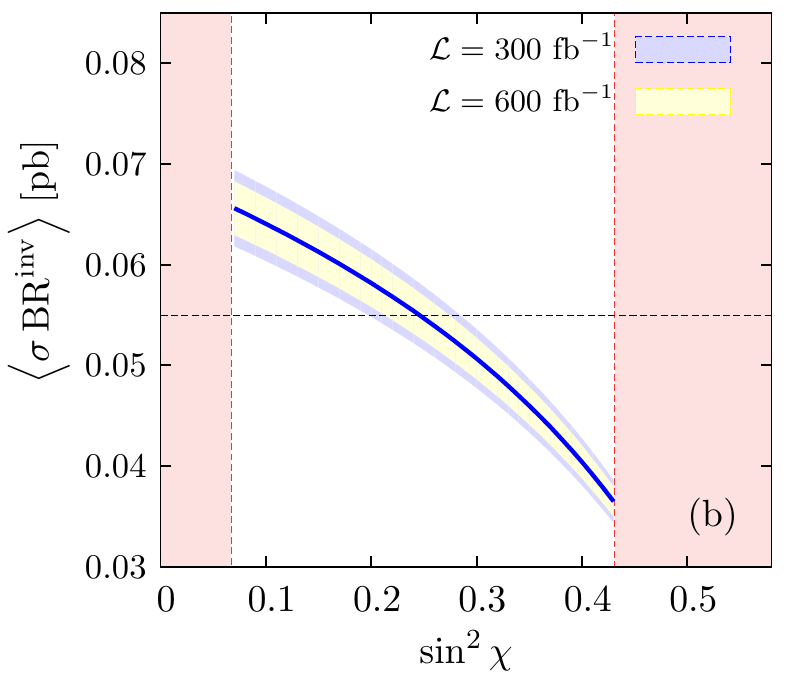} \\
    \includegraphics[height=0.30\textwidth]{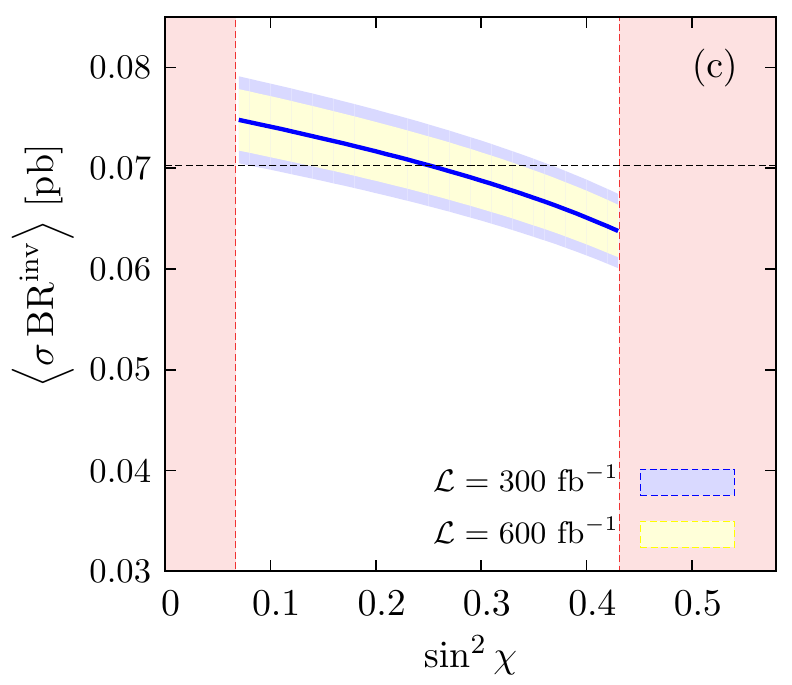}
    \hspace*{0.15\textwidth}
    \parbox{0.30\textwidth}{ \vspace{-6.cm} \caption{\label{fig:avBR}
        \it (a) $\left\langle \sigma \bri \right\rangle$ corresponding
        to reference point \#1.  (cf. Tabs.~\ref{tab:sinex} and
        \ref{tab:lagextract}).  Both Higgs bosons exhibit
        significances of $3\sigma$ or more in the visible sector for
        ${\cal{L}}=30~\ifb$ and will be well-established with
        ${\cal{L}}=300,~600~\ifb$. The dotted line gives the true
        Monte Carlo value. (b) and (c) Extracting the mixing
        parameters in the reference points \#2 and \#3,
        respectively.}}
\end{figure}
\begin{table}[!b]
\begin{tabular}{l|cc|ccc|ccc}
  \hline
  & \multicolumn{2}{c|}{reference point \#1}    & \multicolumn{3}{c|}{reference point \#2}     & \multicolumn{3}{c}{reference point \#3}      \\
  \hline
  $v_s$ [GeV]             &  \multicolumn{2}{c|}{$246.22$ }& \multicolumn{3}{c|}{$246.22$ }&     \multicolumn{3}{c}{$246.22$ }       \\ 
  $\lambda_s$          & $0.18\pm 0.01$  & [0.19]  & $0.58\pm 0.03$ & ($0.58\pm 0.02$) &  [0.58] & $1.04\pm 0.18$ & ($1.03\pm 0.12$) &   [1.03]\\
  $v_h$ [GeV]            & $85.72\pm 32.88$ & [85.75]   &     $36.19\pm 5.06$  & ($36.42\pm 3.63$) &   [36.42]   &       $55.03\pm 27.35$  &   ($58.11\pm 18.94$) & [60.28]     \\
  $\lambda_h$         & $1.53 \pm 0.10$   &  [1.52]            &                    $12.21 \pm 1.25$ & ($12.12 \pm 0.89$) & [12.11]  &  $7.61\pm 3.51$  & ($7.19\pm 2.23$) &  [6.97]\\
  $|\eta_\chi|$          &      $0.13\pm 0.40$ &  [0.13]   &         $3.67\pm 0.53$  &  ($3.66\pm 0.38$)   & [3.61]&       $4.52\pm 2.23$ &  ($4.38 \pm 1.50$) & [4.40]   \\
  \hline
\end{tabular}
\caption{\label{tab:lagextract} \it Extracted Lagrangian parameters of
  the standard sector, the hidden sector, and their coupling for
  ${\cal{L}}=300~\ifb~(600~\ifb)$. Not included are uncertainties that
  arise from measurements of the masses $W$, $H_1$, and $H_2$ and the
  coupling $\Lambda_{211}$. The values in squared brackets are the
  true input.}
\end{table}

Quantitatively, the determination of the mixing angle and the
separation of the two invisibly decaying Higgs states can vary
significantly.  For the reference point \#1 the two Higgs masses are
close to each other, so the SM production cross sections are almost
the same and the sum of both invisible production and decay rates is
nearly independent of the mixing parameter. This is reflected in the
almost horizontal curves in Fig.~\ref{fig:avBR}(a).  In addition, the
measurement of the self-coupling is very difficult as the decay of the
light $H_2$ to two on-shell $H_1$ particles is kinematically
forbidden.  On the other hand, for sufficiently large $H_1,H_2$ mass
splittings and away from maximal mixing
$\sin^2\chi\approx\cos^2\chi\approx 0.5$, as realized in the reference
points \#2 and \#3, these shortcomings are largely removed\footnote{
    For the light Higgs mass chosen in this study, we assume the
    cascade $H_2\rightarrow H_1 H_1$ to be reconstructed from the
    $H_1\rightarrow \gamma\gamma$, which, however, has a smaller
    branching ratio than the competing channel $H_1\rightarrow b\bar
    b$ but similar significance \cite{prod12}. The subjet
    analysis as considered in Ref.~\cite{Butterworth:2008iy} could,
    eventually, allow the reconstruction $H_2\rightarrow H_1H_1\rightarrow b\bar b
    b\bar b, 4g, \tau \bar \tau \tau \bar \tau$. It should be noted
    that a large fraction of $H_2$ Higgs bosons decay to $H_1 H_1$ pairs,
    facilitating the measurement of the $H_2 H_1 H_1$ coupling compared
    with the measurement of the Higgs trilinear coupling in the Standard
    Model.}: since the
SM cross section $\sigma_2^\text{SM}$ is non-dominant, the cross
section for invisible Higgs decays does depend on the mixing
parameter. As a result, we can determine $\sin^2 \chi$, as shown in
Fig.~\ref{fig:avBR}(b,c). In addition, on-shell cascade decays
$H_2\rightarrow H_1H_1$ open and can be exploited to determine
$\Lambda_{211}$. In such cases, Tab.~\ref{tab:lagextract} shows that
the Higgs potential parameters of the Higgs portal can be determined
{\it in toto}: The extraction of the parameters improves upon
increased statistics.

\section{Summary}

A two-state Higgs portal has exciting experimental implications for
the LHC.  Depending on the parameters, three scenarios in which no or
one or two Higgs bosons are observed could be realized.
\begin{itemize}
\item[--] Even though two Higgs bosons are incorporated in the system,
  for medium size mixing none would be observed. This leads to the
  startling conclusion that non-observation of Higgs bosons, for
  luminosities typical for the discovery of the standard Higgs boson,
  does not imply that the Higgs mechanism is falsified.
\item[--] If one Higgs boson is observed experimentally, the couplings
  must be investigated to conclude that the standard mechanism is
  realized. Mixing of the states in the Higgs portal predicts the
  uniform reduction of the couplings to the particles in the Standard
  Model. Enhanced by decays into the hidden sector, the decay
  properties and the production rates are affected, resulting in the
  universal reduction of the twin-width ratios of partial widths over
  the total width which can be observed directly at LHC.
\item[--] Finally, if two Higgs bosons are detected, subsequently
to the profile of the particles, masses and mixing, the microscopic
elements of the Higgs portal can in principle be reconstructed. This
is a difficult experimental task since control over invisible channels
and two-Higgs final states is necessary for drawing a fine-grained
picture. Nevertheless, in this situation we can determine the vacuum
expectation values of the two Higgs fields, the two quartic couplings
of the fields in the standard and the hidden sector as well as the
quartic coupling of the two Higgs fields across the portal.
\end{itemize}
Thus, the Higgs portal including a two-state system offers on one the
hand a simple but interesting extension of the SM Higgs mechanism; on
the other hand, it leads to an exciting lookout to a novel hidden
sector in nature, which can phenomenologically be accessed at the LHC.

\acknowledgments{}
PMZ thanks the Institut f\"ur Theoretische Teilchenphysik und
Kosmologie for the warm hospitality extended to him at RWTH Aachen
University.  DZ und PMZ are grateful to the Institut f\"ur
Theoretische Physik of Heidelberg University for the hospitality
during several visits. Part of the work by DZ is supported 
by the GDR Terascale of the CNRS.


\end{document}